\documentclass{aa}
\usepackage{graphicx}
\usepackage{epsfig}
\usepackage{graphics}

\begin{document}

\title{The proton low-mass microquasar: high-energy emission}
\author{G. E. Romero\inst{1,2,}\thanks{Member of CONICET, Argentina} \and G. S. Vila\inst{1,}\thanks{Fellow of CONICET, Argentina}}
  
\institute{Instituto Argentino de Radioastronom\'{\i}a (CCT La Plata - CONICET), C.C.5, (1984) Villa Elisa, Buenos Aires, Argentina \and Facultad de Ciencias Astron\'omicas y Geof\'{\i}sicas, Universidad Nacional de La Plata, Paseo del Bosque s/n, 1900, La Plata, Argentina}

\offprints{G. S. Vila \\ \email{gvila@iar-conicet.gov.ar}}

\titlerunning{The proton microquasar}

\authorrunning{G.E. Romero and G.S. Vila}

\abstract
{A population of unidentified gamma-ray sources is forming a structure resembling a halo around the Galactic center. These sources are highly variable, and hence they should be associated with compact objects. Microquasars are objects undergoing accretion  with relativistic jets; if such an object has a low-mass, evolved, donor star, it might be found in the Galactic halo. If these low-mass microquasars can generate detectable gamma-ray emission, then they are natural candidates to account for the halo high-energy sources.}
{We aim to construct models for high-energy emission of low-mass microquasars, which could produce a significant luminosity in the gamma-ray domain.}
{We consider that a significant fraction of the relativistic particles in the jets of low-mass microquasars are protons and then we study the production of high-energy emission through proton synchrotron radiation and photopion production. Photopair production and leptonic processes are considered as well. We compute a number of specific models with different parameters to explore the possibilities of this scenario.}
{We find that important luminosities, in the range of $10^{34}-10^{37}$ erg s$^{-1}$, can be achieved by proton synchrotron radiation in the Gamma-Ray Large Area Space Telescope (GLAST) energy range, and lower, but still significant luminosities at higher energies for some models.}
{We conclude that the {proton microquasar model} offers a very interesting alternative to account for the halo gamma-ray sources and presents a variety of predictions that might be tested in the near future by instruments like GLAST, the High-Energy Stereoscopic System II (HESS II), the Major Atmospheric Gamma-ray Imaging Cherenkov telescope II (MAGIC II), and neutrino telescopes like IceCube. }
 
 
\maketitle
 
\section{Introduction}

Microquasars (MQs) are accreting X-ray binaries that produce jets of relativistic particles (e.g., Mirabel \& Rodr\'{\i}guez 1999). According to the type of donor star, microquasars can be classified into high-mass and low-mass microquasars (HMMQs and LMMQs, respectively). Soon after their discovery by Mirabel et al. (1992), it was proposed that MQs might be sources of gamma rays, as it is the case for some active galactic nuclei, in particular, flat-spectrum radio quasars and BL Lacertae (BL Lac) objects (Levinson \& Blandford 1996). A variety of models were then developed for HMMQs, mostly based on interactions between relativistic particles in the jets  (both leptons and hadrons) and matter and photon fields from the hot, massive primary star (see Romero 2005, and references therein; Bosch-Ramon et al. 2006; Bosch-Ramon 2007, and references therein). The primary motivation for these models was to explain the existence of a population of variable and unidentified gamma-ray sources on the galactic plane (Romero 2001; Kaufman Bernad\'o et al. 2002; Bosch-Ramon et al. 2005). These sources are young and mostly located in the inner spiral arms of the Galaxy. Imaging air Cherenkov telescopes have recently detected HMMQs (e.g., Cygnus X-1 and LS I +61 303 by the Major Atmospheric Gamma-ray Imaging Cherenkov telescope, MAGIC, Albert et al. 2006, 2007; and LS 5039 by the High-Energy Stereoscopic System, HESS, Aharonian et al. 2005). There is some controversy around the nature of some of these objects, especially of LS I +61 303; we refer the reader to Romero et al. (2007) for a discussion.

There is a population of variable unidentified sources forming a halo around the Galactic center (Grenier 2001, Nolan et al. 2003). It is expected that this group of sources is old and presents a distribution similar to that of the globular clusters, but with no direct positional correspondence. Since they are extremely variable, it has been suggested that they might be LMMQs (i.e., compact objects) expelled from globular clusters or the galactic plane (Grenier 2001, 2005). 

Different types of models have been proposed to explain gamma-ray emission in LMMQs, where the external fields to the jet itself are exiguous. Aharonian \& Atoyan (1998) and Atoyan \& Aharonian (1999) developed time-dependent synchrotron self-Compton (SSC) models for bursting sources. Bosch-Ramon et al. (2006) explored SSC models for the case of sources with steady inhomogeneous jets. Markoff et al. (2001) presented a synchrotron model with emission reaching the hard X-rays. Grenier et al. (2005), on the other hand, calculated the interaction of a relativistic leptonic jet with the photon fields from the corona, the accretion disk and the dim star, concluding that external inverse Compton interactions cannot be invoked to explain the high gamma-ray luminosities (in the range $10^{34}-10^{36}$ erg s$^{-1}$) inferred from the Energetic Gamma-Ray Experiment Telescope (EGRET) detections.

Hadronic models based on inelastic proton-proton collisions (e.g., Romero et al. 2003) are not expected to work in the bare environment of old and cold stars as those of LMMQs. However, Levinson \& Waxman (2001) proposed that photopion production could be important in MQs if strong synchrotron radiation fields are generated at the base of the jets. These authors developed a simple model to predict the neutrino emission of MQs, where the neutrinos are generated through the decay of charged pions created in proton-photon interactions.  

Here we explore in more detail a proton model for LMMQs, with emphasis on the electromagnetic output. We consider a system where the jets have a content in relativistic protons, which are accelerated by shocks in a compact region close to the accreting object. Electrons are accelerated as well in this region. We study models with different ratios of primary relativistic protons to electrons. The jet itself is considered as a mildly relativistic macroscopic fluid launched by some type of magneto-centrifugal mechanism, and hence it is initially assumed to be in equipartition with the magnetic field (as considered, e.g., by Bosch-Ramon et al. 2006). The jet power is taken as a fraction of the accretion power. In the case of advective accretion flows, the spectral energy distribution will be dominated by the non-thermal emission. Among the processes that cool the relativistic particles we consider synchrotron radiation for both electrons and protons\footnote{The potential of proton synchrotron radiation in the context of AGNs has been explored by Aharonian (2002).}, inverse Compton (IC) interactions with the locally produced fields, photopair and photopion production, and adiabatic losses. The results show that {\sl proton microquasars} might be detected by forthcoming satellites like the Gamma-Ray Large Area Space Telescope (GLAST) and by ground-based air Cherenkov telescopes with a low-energy threshold.

The structure of the paper is as follows. In Sect. 2 we outline the basics of the model and we introduce all the relevant radiative processes. In Section 3 we present the results of our calculations for a number of specific models. Section 4 presents the effects of internal photon-photon absorption on the production spectra. Section 5 contains a discussion of the results. We close in Sect. 6 with a brief summary and the prospects of our model for future applications.

\section{Model}

\subsection{Basic scenario}

\begin{figure}[!h]
  \centering
  \includegraphics[trim=20 180 90 180,clip,width=0.4\textwidth, keepaspectratio]{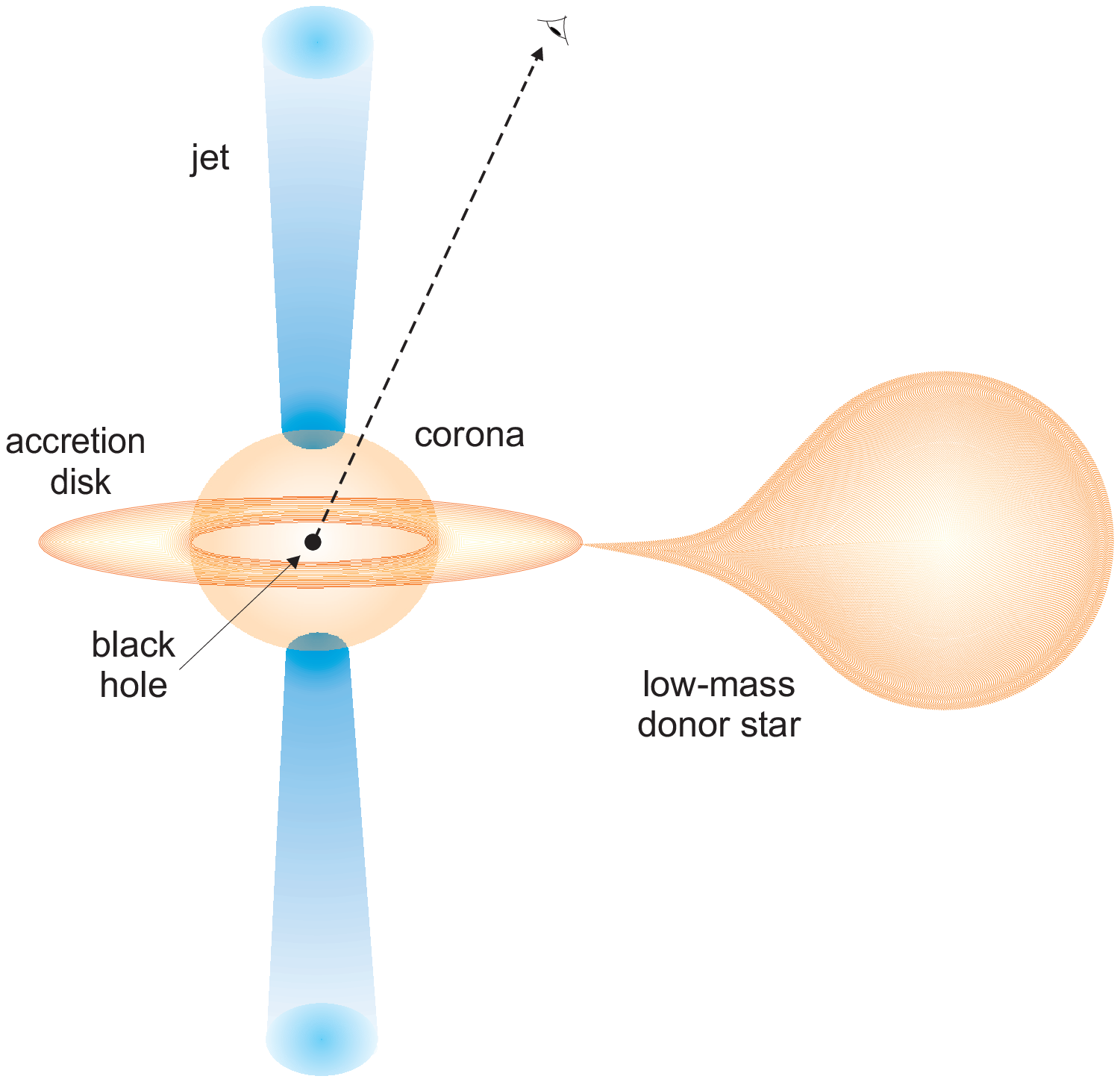}
  \includegraphics[trim=100 300 230 250,clip,width=0.3\textwidth, keepaspectratio]{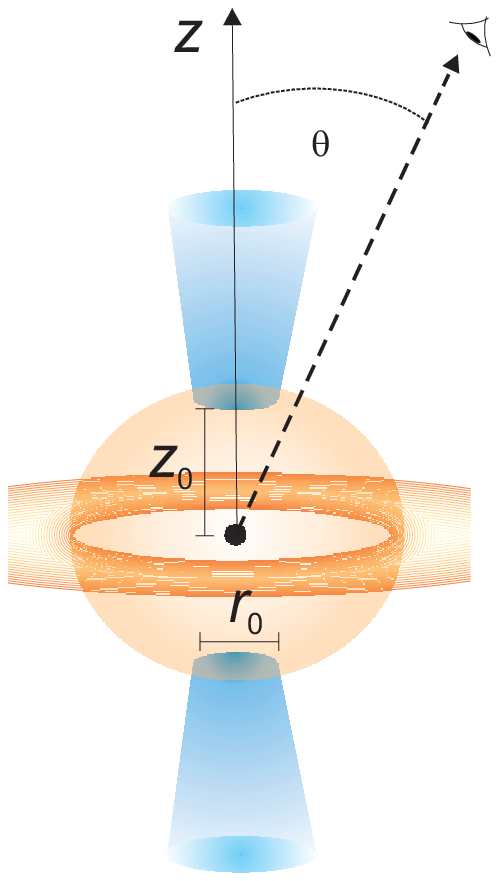} 
  \caption{Top panel: Components of a low-mass microquasar. Bottom panel: Detail of the jet launching region with the relevant geometric parameters.}
  \label{fig:mqscheme}
\end{figure}

The low-hard state of X-ray binaries is characterized by the presence of steady jets and a hot corona around the compact object (see Fig. \ref{fig:mqscheme}). We assume that the jets are launched perpendicularly to the orbital plane of the binary, at a distance $z_0$ from the compact object and with an initial radius $r_0=0.1\,z_0$. Afterwards the outflow expands as a cone of radius $r(z)=r_0\left(z/z_0\right)$.  Following the disk-jet coupling hypothesis of Falcke \& Biermann (1995), the kinetic power of each jet is assumed to be proportional to the accretion power, $L_{\rm{jet}}=q_{\rm{jet}} L_{\rm{accr}}$, with $q_{\rm{jet}}=0.1$ (Falcke \& Biermann 1995; K\"ording et al. 2006). A fraction $L_{\rm{rel}}=q_{\rm{rel}}  L_{\rm{jet}}$ of this power is in the form of relativistic particles. We include both  hadronic and  leptonic content, $L_{\rm{rel}}=L_{p}+L_{e}$. The way in which energy is divided between hadrons and leptons is unknown, but different scenarios can be taken into account by setting $L_{p}=aL_{e}$. In order to have a proton-dominated jet, we consider models with $a>1$, as well as the case of equipartition between hadrons and leptons, $a=1$.

\vspace{0.2cm}

The magnetic field in the jets decreases with the distance $z$ to the black hole\footnote{MQs with a weakly-magnetized neutron star are also possible, e.g. Cir X-1 and Sco X-1; see Migliari \& Fender (2006) for a discussion.}. If the flow expands adiabatically, $B(z)$ is given by 

\begin{equation}
	B(z)=B_0\left(\frac{z_0}{z}\right).
	\label{Bfield1}
\end{equation}
 
\noindent The value of $B_0=B\left(z_0\right)$ can be estimated by equipartition between magnetic and kinetic energy densities at $z_0$,

\begin{equation}
	\frac{B_0^2}{8\pi}=\frac{L_{\rm{jet}}}{2\pi r_0^2 v_{\rm{jet}}}.
	\label{Bfield2}
\end{equation}  

\noindent Here $v_{\rm{jet}}$ is the bulk velocity of the jets. We find $B_0\approx2\times10^7$ G for a jet with  parameters similar to those of the well-known LMMQ XTE J1118+480, see Table \ref{table}.

\begin{table}[!h]
    \caption[]{Model parameters}
\begin{tabular}{ll}
\hline\noalign{\smallskip}
Parameter & Value\\[0.01cm]
\hline\noalign{\smallskip}
$L_{\rm{accr}}$: accretion power [erg s$^{-1}$] & $1.7\times10^{39}$$^{(1)}$ \\[0.01cm]
$q_{\rm{jet}} $: disk-jet coupling constant & 0.1 \\[0.01cm]
$q_{\rm{rel}} $: jet's content of relativistic particles & 0.1\\[0.01cm]
$a$: hadron-to-lepton energy ratio & $1$ to $10^3$\\[0.01cm]
$z_0$: jet's launching point [cm] & $10^8$$^{(2)}$\\[0.01cm]
$z_{\rm{max}}$: extent of acceleration region & $5\,z_0$\\[0.01cm]
$\Gamma_{\rm{jet}}$: jet's bulk Lorentz factor & 1.5$^{(1)}$ \\[0.01cm]
$B_0$: magnetic field at base of jet [G]& $2\times 10^7$ \\[0.01cm]
$\alpha$: particle injection spectral index & $1.5 - 2.2$\\[0.01cm]
$\eta$: acceleration efficiency & $10^{-4} - 0.1$\\[0.01cm]
$E^{\rm{min}}$: minimum particle energy & $2 - 100$ $mc^2$\\[0.01cm]
$\theta$: viewing angle & $30^\circ$\\[0.01cm]
\hline\\[0.005cm]
\multicolumn{2}{l}{
$^{(1)}$ Typical value for the LMMQ XTE J1118+480.}  \\[0.01cm]
\multicolumn{2}{l}{
$^{(2)}$ $50 R_{\rm{Schw}}$ for XTE J1118+480 ($8\,M_{\odot}$ black hole).}  \\[0.01cm]
\end{tabular}	
  \label{table}
\end{table}

\vspace{0.2cm}

In our model, particle acceleration is constrained to take place in a compact region near the base of the jets, extending up to a distance $z_{\rm{max}}$ of a few times $z_0$. We assume an injection function that is a power-law in the energy of the particles,

\begin{equation}
	Q\left(E,z\right)=Q_0\frac{E^{-\alpha}}{z} \qquad [Q]=\rm{erg}^{-1}\rm{s}^{-1}\rm{cm}^{-3}.
\label{injection}
\end{equation}  

\noindent This type of spectral distribution arises naturally in astrophysical environments as a consequence of diffusive particle acceleration by shock waves. Both the cases of  hard and  soft particle injection were considered in our calculations. For these spectra we adopted $\alpha=1.5$ and $\alpha=2.2$, respectively. Also, we added an explicit $z$-dependence to reduce the injection rate from the distance to the compact object. The normalization constant $Q_0$ for each type of particle was obtained from $L_{e,p}$ as

\begin{equation}
 L_{e,p}=\int_{V}\mathrm{d}^3r\int_{E_{e,p}^{\rm{min}}}^{E_{e,p}^{\rm{max}}\left(z\right)}\mathrm{d}E_{e,p}\,E_{e,p}\,Q_{e,p}(E_{e,p},z).
\label{norminj}
\end{equation}    

\begin{figure*}[!t]
  \centering
  \includegraphics[trim=10 10 0 10,clip,width=0.41\textwidth, keepaspectratio]{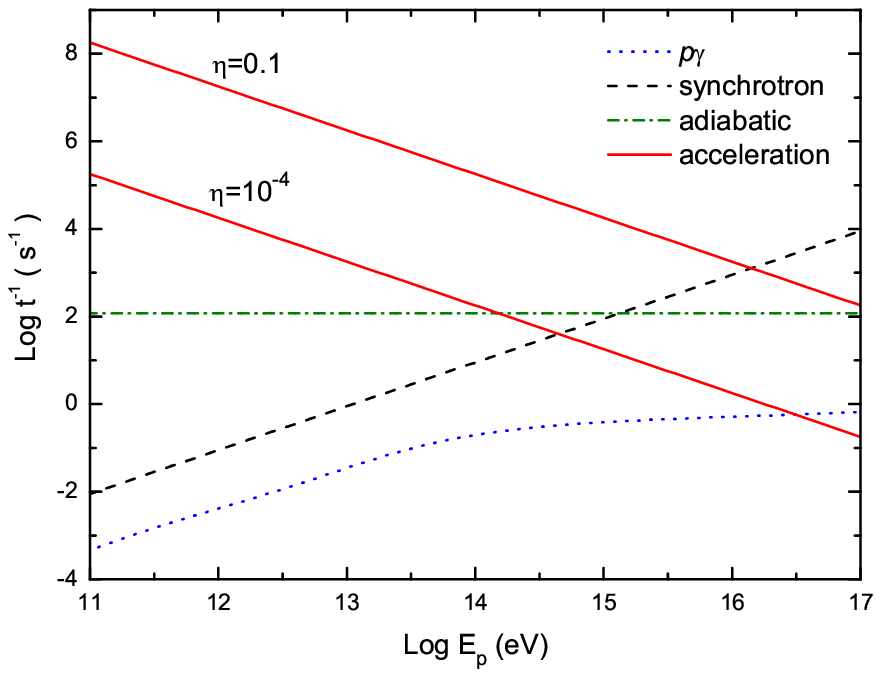}
  \includegraphics[trim=0 10 20 10,clip,width=0.4\textwidth, keepaspectratio]{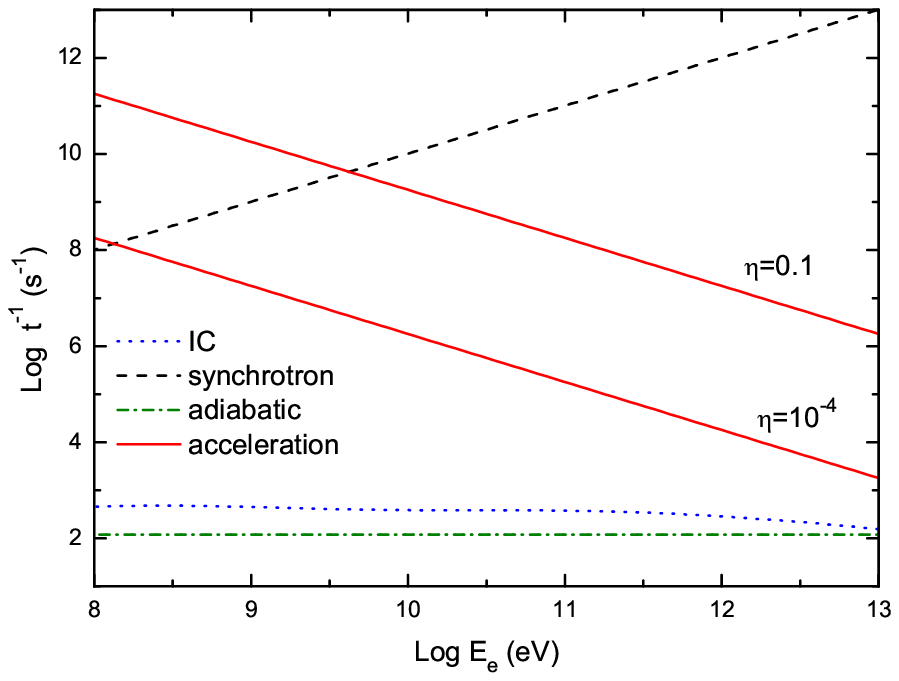} 
  \caption{Acceleration and cooling rates at the base of the jet for protons (left panel) and electrons (right panel), calculated for representative values of the model parameters ($E^{\rm{min}}=100\,mc^2$, $a=1$ and $\alpha=2.2$). }
  \label{fig:cool}
\end{figure*}

\noindent The maximum energy that a relativistic particle can attain depends on the acceleration mechanism and the different processes of energy loss. The acceleration rate $t^{-1}_{\rm{acc}}=E^{-1}dE/dt$ for a particle of energy $E$ in a magnetic field $B$ is given by

\begin{equation}
	t^{-1}_{\rm{acc}}=\frac{\eta ecB}{E},
	\label{accrate}
\end{equation}
   
\noindent where $\eta\leq1$ is a parameter that characterizes the efficiency of the acceleration. Two cases were considered: $\eta=0.1$ and $\eta=10^{-4}$, describing an efficient and a poor acceleration, respectively. In Fig. \ref{fig:cool}, we show the cooling rates at $z_0$ for different processes of energy loss, together with the acceleration rate, for protons and leptons. Under the physical conditions of our model the main channel of energy loss for electrons is synchrotron radiation and we can approximate $t^{-1}_{\rm{loss}}\approx t^{-1}_{\rm{synchr}}$; for protons both synchrotron and adiabatic losses are relevant, $t^{-1}_{\rm{loss}}\approx t^{-1}_{\rm{synchr}}+t^{-1}_{\rm{ad}}$. The synchrotron cooling rate for a particle of mass $m$, charge $e$, and energy $E$ in a region of magnetic energy density $U_B$ is

\begin{equation}
	t^{-1}_{\rm{synchr}}=\frac{4}{3}\left(\frac{m_e}{m}\right)^3\frac{c\sigma_TU_B}{m_ec^2}\frac{E}{mc^2},
	\label{coolratesy}
\end{equation}  

\noindent whereas

\begin{equation}
	t^{-1}_{\rm{ad}}=\frac{2}{3}\frac{v_{\rm{jet}}}{z}.
	\label{coolratead}
\end{equation} 
\noindent The maximum energy is thus fixed by requiring  

\begin{equation}
	t^{-1}_{\rm{acc}}(E,z)=t^{-1}_{\rm{loss}}(E,z).
	\label{maxenergcond}
\end{equation}

\begin{figure*}[!t]
  \centering
  \includegraphics[trim=10 10 0 10,clip,width=0.435\textwidth, keepaspectratio]{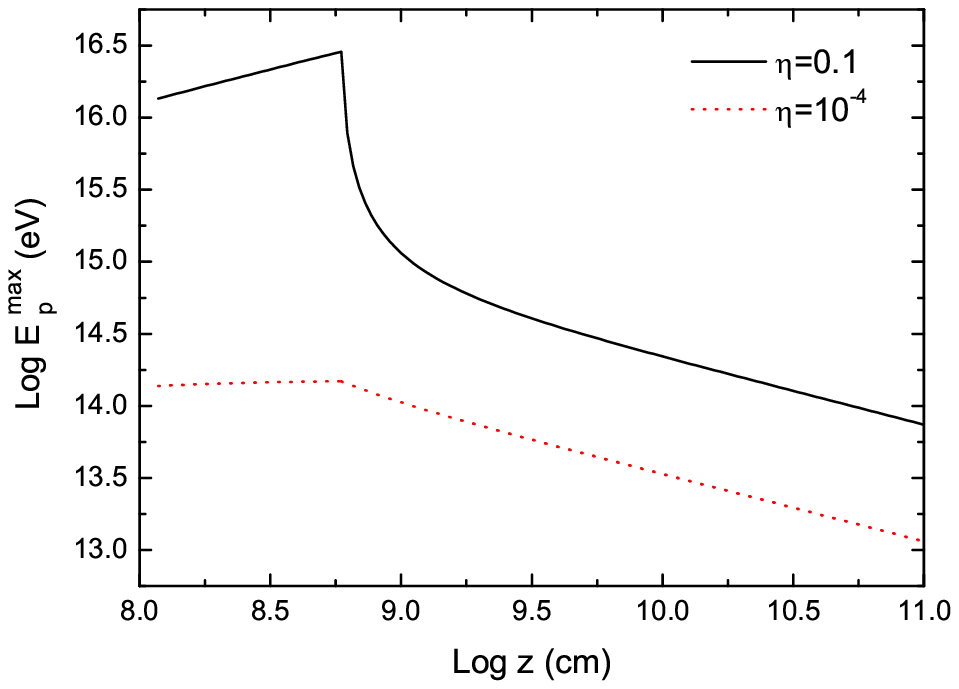}
  \includegraphics[trim=0 10 20 10,clip,width=0.42\textwidth, keepaspectratio]{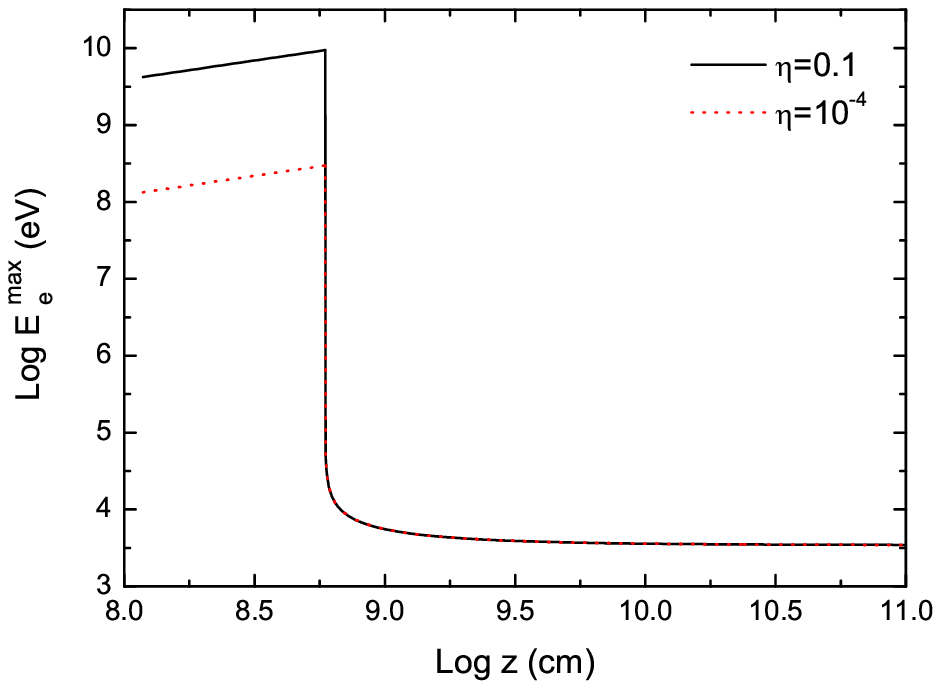} 
  \caption{Evolution of the maximum (kinetic) energy of protons (left panel) and electrons (right panel) with the distance $z$ to the compact object. Outside the acceleration region particles cool rapidly due to adiabatic and synchrotron losses.}
  \label{fig:emax}
\end{figure*}

\noindent For electrons this yields

\begin{equation}
	E_{e}^{\rm{max}}(z)\approx1.3\times10^{10}\left(\eta\frac{z}{z_0}\right)^{1/2}\,\rm{eV},
	\label{maxleptons}
\end{equation}  

\noindent and for protons

\begin{equation}
	E_{p}^{\rm{max}}(z)\approx4.5\times10^{16}\left(\eta\frac{z}{z_0}\right)^{1/2}\,\rm{eV}.
	\label{maxprotons}
\end{equation}  

\noindent Particles cool rapidly after leaving the acceleration region, as can be seen from Fig. \ref{fig:emax}, which shows the evolution of $E^{\rm{max}}(z)$. The size of the acceleration region can impose constraints on the maximum energy; this occurs only in the case in which the particle giroradius $r_{\rm{g}}(z)=E/eB(z)$ is greater than $r_{\rm{jet}}(z)$. Nevertheless, in our model maximum energies are in all cases given by Eqs. (\ref{maxleptons}) and (\ref{maxprotons}).

We have calculated the steady state particle distributions $N(E,z)$ in the ``one-zone'' approximation (Khangulyan et al. 2007). This approximation is valid if the losses are very strong in the acceleration region and diffusion can be neglected. Then the transport equation (Ginzburg \& Syrovatskii 1964) can be written as

\begin{equation}
	\frac{\partial}{\partial E}\left[\left.\frac{dE}{dt}\right|_{\rm{loss}}N(E,z)\right]+\frac{N(E,z)}{t_{\rm{esc}}}=Q(E,z).
	\label{transpeq1}
\end{equation} 

\noindent Here $t_{\rm{esc}}$ is the particle escape time from the acceleration region, which we estimated as $t_{\rm{esc}}\approx z_{\rm{max}}/v_{\rm{jet}}$. This equation includes energy losses, removal of particles, and injection. It provides a one-zone description of the flow since, for each  $z$, it neglects the contribution to $N\left(E,z\right)$ of particles coming from other regions in the jet. It remains a good approximation as long as the acceleration/emission region is compact (see Khangulyan et al. 2008 for a different approach). This is better fulfilled in the case of leptons, since they cool almost completely near the base of the jet where the magnetic field is stronger. The exact analytical solution to Eq. (\ref{transpeq1}) can be found in Khangulyan et al. (2007):

\begin{eqnarray}
N(E,z)=\left|\frac{dE}{dt}\right|_{\rm{loss}}^{-1}\int_E^{E^{\rm{max}}\left(z\right)}&&\mathrm{d}E^\prime  Q(E^\prime,z)\nonumber \\ && \times\exp\left(-\frac{\tau(E, E^\prime)}{t_{\rm{esc}}}\right),
	\label{transpeq2}
\end{eqnarray} 

\noindent where

\begin{equation}
	\tau(E, E^\prime)=\int_E^{E^\prime}\mathrm{d}E^{\prime\prime}\left|\frac{dE^{\prime\prime}}{dt}\right|_{\rm{loss}}^{-1}.
	\label{transpeq3}
\end{equation} 
 
\noindent The resulting distributions have a power-law dependence on the energy of the particles. In the case of protons, $N_p(E_p,z)$ mimics the injection function, $N_p(E_p,z)\propto E_p^{-\alpha}$. The synchrotron cooling timescale is much shorter for electrons and they radiate a significant fraction of their energy before escaping. This modifies the injection spectrum making it softer, and $N_e(E_e,z)$ can be well approximated as $N_e(E_e,z)\propto E_e^{-(\alpha+1)}$. 

\subsection{Radiative processes}

We consider three processes of interaction of relativistic particles with the fields in the jet: synchrotron radiation of protons and electrons, proton-photon inelastic collisions ($p\gamma$) and IC scattering. 

To calculate the synchrotron spectra we use the classical formulae, see for example Blumenthal \& Gould (1970). In the jet co-moving reference frame, where particle distributions are isotropic, the power radiated by a single particle of energy $E$, and pitch angle $\alpha$ is given by 

\begin{eqnarray}
P_{\rm{synchr}}\left(E_\gamma,E,z,\alpha\right)=\frac{\sqrt{3}e^3B(z)}{4\pi mc^2h}\frac{E_\gamma}{E_{\rm{c}}}\int_{E_\gamma/E_{\rm{c}}}^\infty \mathrm{d}\xi\,K_{5/3}(\zeta),&&\nonumber\\
&&
	\label{sypower}
\end{eqnarray} 

\noindent where $K_{5/3}(\zeta)$ is a modified Besell function. The synchrotron power peaks sharply around the characteristic energy $E_{\rm{c}}$, given by

\begin{equation}
E_{\rm{c}}=\frac{3heB(z)\sin\alpha}{4\pi mc}\left(\frac{E}{mc^2}\right)^2.
	\label{sychar}
\end{equation}

\noindent Denoting the volume of the emission region by $V$, the total luminosity is 

\setlength\arraycolsep{1pt}

\begin{eqnarray}
L_{\rm{synchr}}\left(E_\gamma\right)=E_\gamma &&\int_V \mathrm{d}^3r\int_{\Omega_\alpha}\mathrm{d}\Omega_\alpha\sin\alpha\nonumber\\
&&\,\,\times\int_{E^{\rm{min}}}^{E^{\rm{max}}\left(z\right)}\mathrm{d}E\,N(E,z)\,P_{\rm{synchr}}.  
	\label{syluminosity}
\end{eqnarray}    

Electrons in the jet interact with the synchrotron radiation fields through IC scattering. As leptons are highly relativistic they can lose an important fraction of their energy in each collision and the Klein-Nishina cross section must be used in the calculations. According to Blumenthal \& Gould (1970), the spectrum of photons scattered by an electron of energy $E_{\rm{e}}=\gamma_{\rm{e}}m_{\rm{e}}c^2$ in a target radiation field of density $n(\epsilon,z)$ is

\begin{equation}
P_{\rm{IC}}\left(E_\gamma,E_{\rm e},\epsilon,z\right)=\frac{3\sigma_{\rm{T}}c}{4\gamma_{\rm{e}}^2}\frac{n\left(\epsilon,z\right)}{\epsilon}F\left(q\right).
	\label{ICpower}
\end{equation} 

\noindent Here $\sigma_{\rm{T}}$ is the Thomson cross section and the function $F\left(q\right)$ is given by

\begin{eqnarray}
F\left(q\right)=2q\ln q+\left(1+2q\right)&&\left(1-q\right)\nonumber\\&&+\frac{1}{2}\left(1-q\right)\frac{\left(q\Gamma_{\rm{e}}\right)^2}{\left(1+\Gamma_{\rm{e}}q\right)},
	\label{Ffunction}
\end{eqnarray} 

\noindent where 

\begin{equation}
\Gamma_{\rm{e}}=\frac{4\epsilon\gamma_{\rm{e}}}{m_{\rm{e}}c^2}
	\label{Gammae}
\end{equation} 

\noindent and

\begin{equation}
q=\frac{E_\gamma}{\Gamma_{\rm{e}}E_{\rm{e}}\left(1-E_\gamma/E_{\rm{e}}\right)}.
	\label{q}
\end{equation}  

\noindent The limit $\Gamma_{\rm{e}}\ll1$ corresponds to Thomson scattering, but Eq. (\ref{ICpower}) is valid for all $\Gamma_{\rm{e}}$, even deep into the Klein-Nishina regime, as long as $\gamma_{\rm{e}}\gg1$. The allowed range of energies for the scattered photons is

\begin{equation}
\epsilon\leq E_\gamma\leq \frac{\Gamma_{\rm{e}}}{1+\Gamma_{\rm{e}}}E_{\rm{e}}.
	\label{allowedE}
\end{equation} 

\noindent Given the electron distribution $N\left(E_{\rm{e}},z\right)$, the total IC spectrum is calculated as

\begin{eqnarray}
L_{\rm{IC}}\left(E_\gamma\right)=E^{2}_\gamma \int_V \mathrm{d}^3r\int_{E_{\rm{e}}^{\rm{min}}}^{E_{\rm{e}}^{\rm{max}}\left(z\right)}\mathrm{d}E_{\rm{e}}&&\,N\left(E_{\rm{e}},z\right)\nonumber\\&&\times\int_{\epsilon_{\rm{min}}}^{\epsilon_{\rm{max}}}\mathrm{d}\epsilon\,P_{\rm{IC}}.
	\label{ICluminosity}
\end{eqnarray} 

\noindent In our case, the scattered photons are those of the synchrotron radiation fields of both protons and electrons, 

\begin{equation}
n\left(\epsilon,z\right)=n_{\rm{synchr}}\left(\epsilon,z\right).
	\label{sydensity}
\end{equation} 
 
\noindent We estimated $n_{\rm{synchr}}$ in the \emph{local approximation} of Ghisellini et al. (1985),

\begin{equation}
n_{\rm{synchr}}\left(\epsilon,z\right)\approx\frac{\varepsilon_{\rm{sy}}\left(\epsilon,z\right)}{\epsilon}\frac{r(z)}{c},
	\label{sylocal}
\end{equation}  

\noindent where $\varepsilon_{\rm{synchr}}\left(\epsilon,z\right)$ is the synchrotron power per unit volume per unit energy.\\

Proton-photon ($p\gamma$) interaction has two main channels: photopair production

\begin{equation}
p+\gamma\rightarrow p+e^-+e^+,
	\label{photopair}
\end{equation}  

\noindent with a threshold energy of $\sim1$ MeV for the photon in the proton rest frame, and photomeson production, which proceeds through two main branches: 

\begin{equation}
p+\gamma\rightarrow p+a\pi^0+b\left(\pi^++\pi^-\right)
	\label{photomeson1}
\end{equation}  

\noindent and

\begin{equation}
p+\gamma\rightarrow n+\pi^++a\pi^0+b\left(\pi^++\pi^-\right),
	\label{photomeson2}
\end{equation}  

\noindent with approximately the same cross section and a threshold energy of $\sim145$ MeV. Here $a$ and $b$ are the pion multiplicities. Charged pions decay in leptons and neutrinos,
 
\begin{equation}
\pi^+\rightarrow\mu^++\nu_\mu, \quad \mu^+\rightarrow e^++\nu_{\rm{e}}+\overline{\nu}_\mu
	\label{piondecay1}	
\end{equation}

\begin{equation}
\pi^-\rightarrow\mu^-+\overline{\nu}_\mu, \quad \mu^-\rightarrow e^-+\overline{\nu}_{\rm{e}}+\nu_\mu,
	\label{piondecay2}	
\end{equation}
 
\noindent whereas neutral pion decay yields photons

\begin{equation}
\pi^0\rightarrow2\gamma.
	\label{piondecay3}	
\end{equation}

\noindent The $p\gamma$ cooling rate for a proton of Lorentz factor $\gamma_p$ in a photon field of density $n\left(\epsilon\right)$ is 
 
\begin{eqnarray}
t^{-1}_{p\gamma, \,i}(\gamma_p)=\frac{c}{2\gamma_p^2}&& \int_{\frac{\epsilon_{\rm{th}}}{2\gamma_p}}^\infty \mathrm{d}\epsilon  \frac{n\left(\epsilon\right)}{\epsilon^2}\nonumber\\&& \times\int_{\epsilon_{\rm{th}}}^{2\epsilon\gamma_p}\mathrm{d}\epsilon^\prime\sigma_{p\gamma,\,i}\left(\epsilon^\prime\right)K_{p\gamma,\,i}\left(\epsilon^\prime\right)\epsilon^\prime,
	\label{coolpg}	
\end{eqnarray} 
 
\noindent where $i=e^\pm,\pi$ denotes the interaction channel, $\epsilon^\prime$ is the energy of the photon in the rest frame of the proton, and $\epsilon_{\rm{th}}$ is the photon threshold energy measured in the same frame; $K_{p\gamma}$ is the inelasticity of the process. The corresponding collision rate is given by a similar expression:  

\begin{equation}
\omega_{p\gamma,\,i}(\gamma_p)=\frac{c}{2\gamma_p^2}\int_{\frac{\epsilon_{\rm{th}}}{2\gamma_p}}^\infty\mathrm{d}\epsilon\frac{n\left(\epsilon\right)}{\epsilon^2}\int_{\epsilon_{\rm{th}}}^{2\epsilon\gamma_p}\mathrm{d}\epsilon^\prime\sigma_{p\gamma,\,i}\left(\epsilon^\prime\right)\epsilon^\prime.
	\label{Kpg}	
\end{equation} 
 
\noindent To estimate the spectrum from the decay of $\pi^0$, we followed  Atoyan \& Dermer (2003). In this approach, the cross section $\sigma_{p\gamma,\,\pi}$ and the inelasticity $K_{p\gamma,\,\pi}$ are approximated by step functions 

\begin{displaymath}
\sigma_{p\gamma,\,\pi}(\epsilon^\prime)\approx\left\{\begin{array}{ll}
340\,\mu\mathrm{barn} \quad 200\,\mathrm{MeV}\leq\epsilon^\prime\leq500\,\mathrm{MeV} \\[0.3cm]
120\,\mu\mathrm{barn} \quad \epsilon^\prime>500\,\mathrm{MeV},
\end{array} \right.
\label{sigmapgamma}
\end{displaymath}

\begin{displaymath}
K_{p\gamma,\,\pi}(\epsilon^\prime)\approx\left\{\begin{array}{ll}
0.2 \quad 200\,\mathrm{MeV}\leq\epsilon^\prime\leq500\,\mathrm{MeV} \\[0.3cm]
0.6 \quad \epsilon^\prime>500\,\mathrm{MeV}.
\end{array} \right.
\label{Kpgamma}
\end{displaymath}

\noindent These energy ranges correspond to the single-pion and multiple-pion production channels, respectively.

The spectra are calculated in the $\delta$-functional approximation in the energy of pions and photons. In the single-pion production channel, each $\pi^{0}$ is produced with an energy $\approx K_1E_{\rm{p}}$ and this energy is equally distributed among the products of its decay, yielding  

\begin{equation}
E_\gamma\approx\frac{1}{2}K_1E_{\rm{p}}=0.1E_{\rm{p}}.
	\label{energies2}	
\end{equation}  
 
\noindent In the multiple-pion production channel almost all the energy lost by the proton is equally distributed among three leading pions $\pi^{0}$, $\pi^{+}$, and $\pi^{-}$. The mean energy of each pion is $\approx K_2E_p/3$ and, therefore, the energy of the photons results the same as in the single-pion production channel.

If $p_1$ and $p_2=1-p_1$ are the probabilities of the $p\gamma$ collision taking place through the single-pion and multiple-pion channel, respectively, and $\xi_{pn}\approx0.5$ is the probability of conversion of a proton into a neutron per interaction, the mean number of neutral pions created per collision is

\begin{equation}
n_{\pi^0}=p_1\left(1-\xi_{pn}\right)+p_2.
	\label{meannumber1}	
\end{equation}    
 
\noindent We can define a mean inelasticity

\begin{equation}
\overline{K}_{p\gamma,\,\pi}\left(\gamma_p\right)=t^{-1}_{p\gamma,\,\pi}\left(\gamma_p\right)\omega_{p\gamma,\,\pi}\left(\gamma_p\right),
	\label{meanKpg}	
\end{equation}
  
\noindent then the probabilities $p_{1,2}$ can be calculated from the relation

\begin{equation}
\overline{K}_{p\gamma,\,\pi}\left(E_p\right)=p_1K_1+\left(1-p_1\right)K_2.
	\label{meanK}	
\end{equation}    

\noindent The emissivity of $\pi^0$ in the $\delta$-functional approximation is 

\begin{eqnarray}
Q_{\pi^0}\left(E_{\pi}\right)&=&\int\mathrm{d}E_pN_p\left(E_p\right)\omega_{p\gamma,\,\pi}\left(E_p\right)\nonumber \\ && \qquad \qquad \quad \times n_{\pi^0}\left(E_p\right)\delta\left(E_\pi-0.2E_p\right) \nonumber \\[0.2cm] 
&=&5N_p\left(5E_\pi\right)\omega_{p\gamma,\,\pi}\left(5E_\pi\right)n_{\pi^0}\left(5E_\pi\right).             
	\label{emispi}	
\end{eqnarray}  
 
\noindent Taking into account that each $\pi^0$ gives two photons, the photon emissivity results

\begin{eqnarray}
Q_{\gamma}\left(E_{\gamma}\right)&=&2\int\mathrm{d}E_\pi\, Q_{\pi^0}\left(E_{\pi}\right)\delta\left(E_\gamma-0.5E_\pi\right) \nonumber \\ 
&=&20N_p\left(10E_\gamma\right)\omega_{p\gamma,\,\pi}\left(10E_\gamma\right)n_{\pi^0}\left(10E_\gamma\right).            
	\label{emisgamma}	
\end{eqnarray} 

\noindent The emissivity of secondary pairs can be estimated in the same way. Both in the single-pion and in the multiple-pion channel each charged pion has an energy $\sim 0.2 E_p$. This energy is equally distributed among the products of meson decay, hence the energy of each electron/positron is, on average, $E_{e}\sim0.05E_p$. The mean number of charged pions created in a $p\gamma$ collision is 

\begin{equation}
n_{\pi^\pm}=\xi_{pn}p_1+2p_2,
	\label{meannumber2}	
\end{equation}  

\noindent and since only one lepton is produced in each decay, the emissivity of pairs is given by

\begin{eqnarray}
Q_{e^{\pm}}\left(E_{e^{\pm}}\right)=20 N_p\left(20E_{e^{\pm}}\right)\omega_{p\gamma,\,\pi}\left(20E_{e^{\pm}}\right)&&\nonumber\\[0.1cm]  && \hspace{-1cm} \times n_{\pi^\pm}\left(20E_{e^{\pm}}\right).         	
\label{emispairspg}	
\end{eqnarray} 
 
The second channel of proton-photon interaction is pair production. This process has been studied, for example, by  Chodorowski et al. (1992), M\"ucke et al. (2000) and Mastichiadis et al. (2005). The emissivity of pairs in the $\delta$-functional approximation is calculated as for the previous process. Now the inelasticity can be approximated by its value at the threshold, $K_{p\gamma,\,e^\pm}=2m_e/m_p$ . Therefore,  

\begin{eqnarray}
Q_{e^{\pm}}\left(E_{e^{\pm}}\right)&=&2\int\mathrm{d}E_p N_p\left(E_p\right)\omega_{p\gamma,\,e^\pm}\left(E_p\right)\delta\left(E_{e^{\pm}}-\frac{m_e}{m_p}E_p\right)\nonumber\\
&=&2\frac{m_p}{m_e}N_p\left(\frac{m_p}{m_e}E_{e^{\pm}}\right)\omega_{p\gamma,\,e^\pm}\left(\frac{m_p}{m_e}E_{e^{\pm}}\right),
	\label{emispairs}	
\end{eqnarray}  

\noindent An useful parametrization of the cross section $\sigma_{p\gamma,\,e^\pm}$ can be found in Maximon (1968).
 
All the expressions above are valid in the jet co-moving reference frame, where the particle distributions are isotropic. The spectral energy distributions in the rest frame of the observer can be obtained by applying the appropriate boost. In what follows, we denote by primed and non-primed quantities those measured in the jet co-moving frame and in the observer frame, respectively. Photon energies in both frames are related by the Doppler factor $D$,

\begin{equation}
E_\gamma=DE_\gamma^\prime.
	\label{Dfactor1}	
\end{equation}  

\noindent The factor $D$ depends on the jet's bulk Lorentz factor $\Gamma_{\rm{jet}}$ and on the viewing angle $\theta$. For an approaching jet,

\begin{equation}
D=\frac{1}{\Gamma_{\rm{jet}}\left(1-\beta_{\rm{jet}}\cos\theta\right)}.
	\label{Dfactor2}	
\end{equation} 

\noindent Finally, the luminosity in the observer frame is given by (e.g., Lind \& Blandford 1985)

\begin{equation}
L_\gamma\left(E_\gamma\right)=D^2L^\prime_\gamma\left(E_\gamma^\prime\right).
	\label{Lboost}	
\end{equation} 

\noindent None of our results depend strongly on the value of the viewing angle as long as is it is not very close to zero, and we fixed $\theta=30^\circ$ for calculation purposes. The values of the different parameters of the model used in the calculations are shown in Table \ref{table}.

\section{Results} 

\begin{figure*}[!t]
  \centering
  \includegraphics[trim=10 18 10 0,clip,width=0.45\textwidth, keepaspectratio]{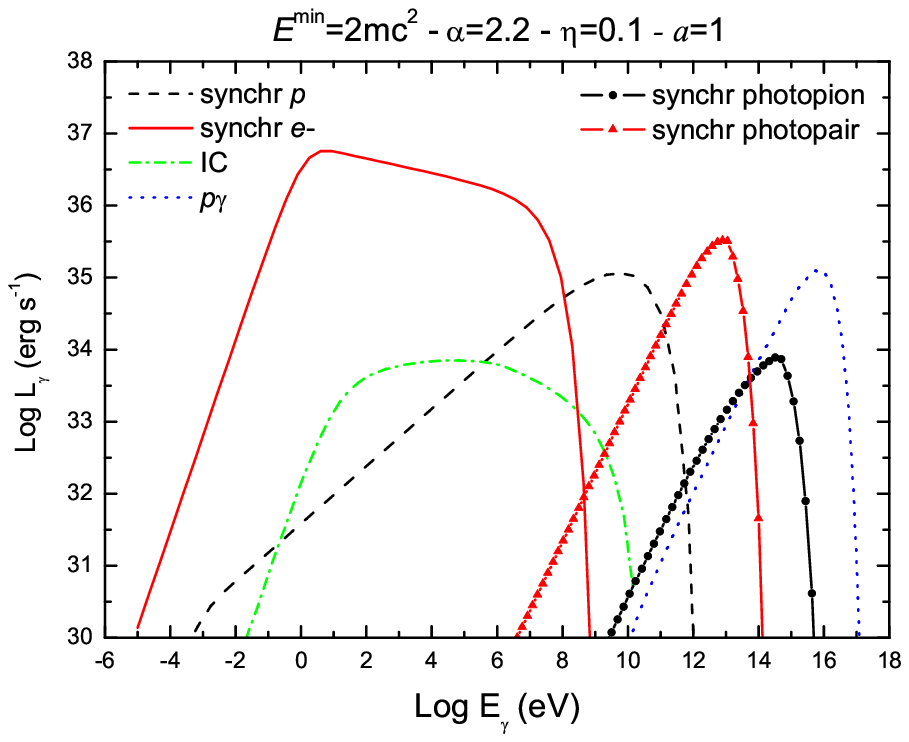}
  \includegraphics[trim=10 18 10 0,clip,width=0.47\textwidth, keepaspectratio]{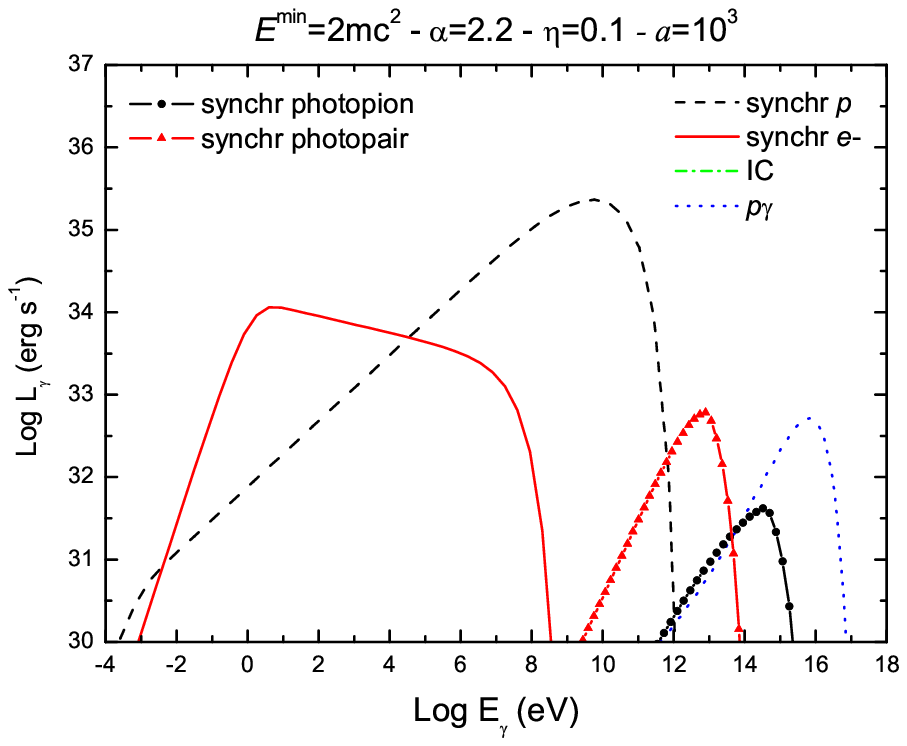} 
  \includegraphics[trim=17 18 5 0,clip,width=0.45\textwidth,keepaspectratio]{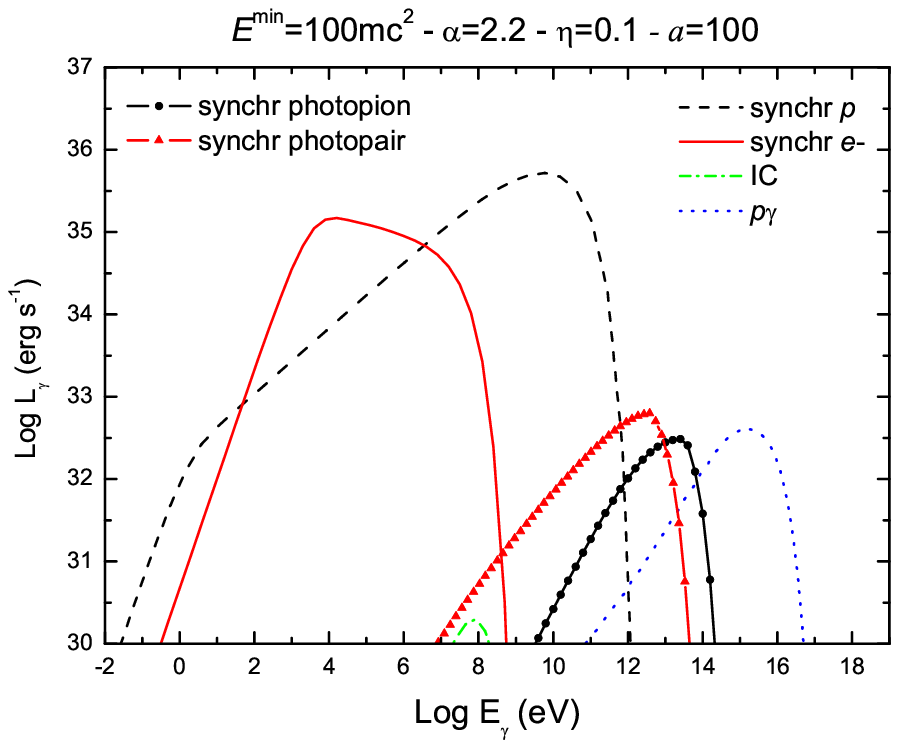} 
  \includegraphics[trim=17 18 5 18,clip,width=0.45\textwidth, keepaspectratio]{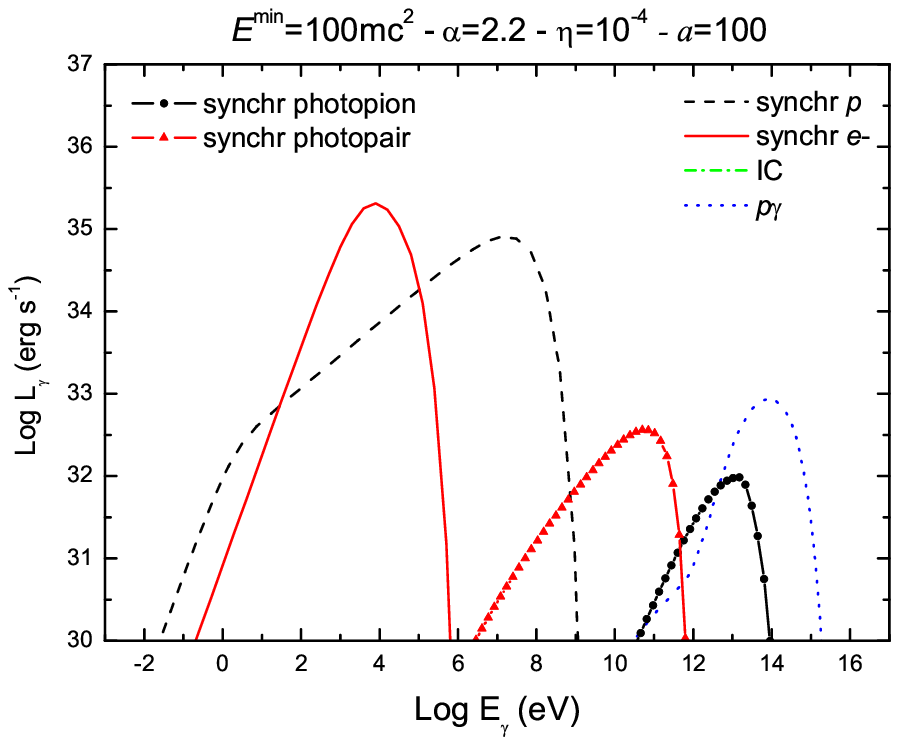}
  \caption{Spectral energy distributions obtained for different values of the parameters in the case of a particle injection index $\alpha=2.2$.}
  \label{fig:seds1}
\end{figure*}

\begin{figure*}[!t]
  \centering
  \includegraphics[trim=10 18 5 0,clip,width=0.45\textwidth, keepaspectratio]{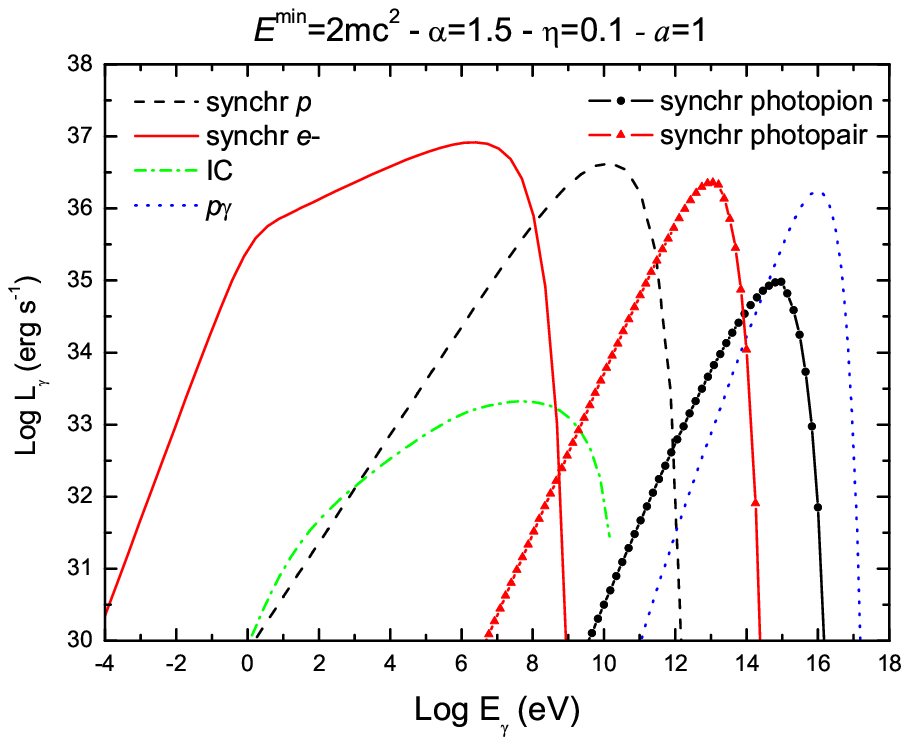}
  \includegraphics[trim=10 18 5 0,clip,width=0.45\textwidth, keepaspectratio]{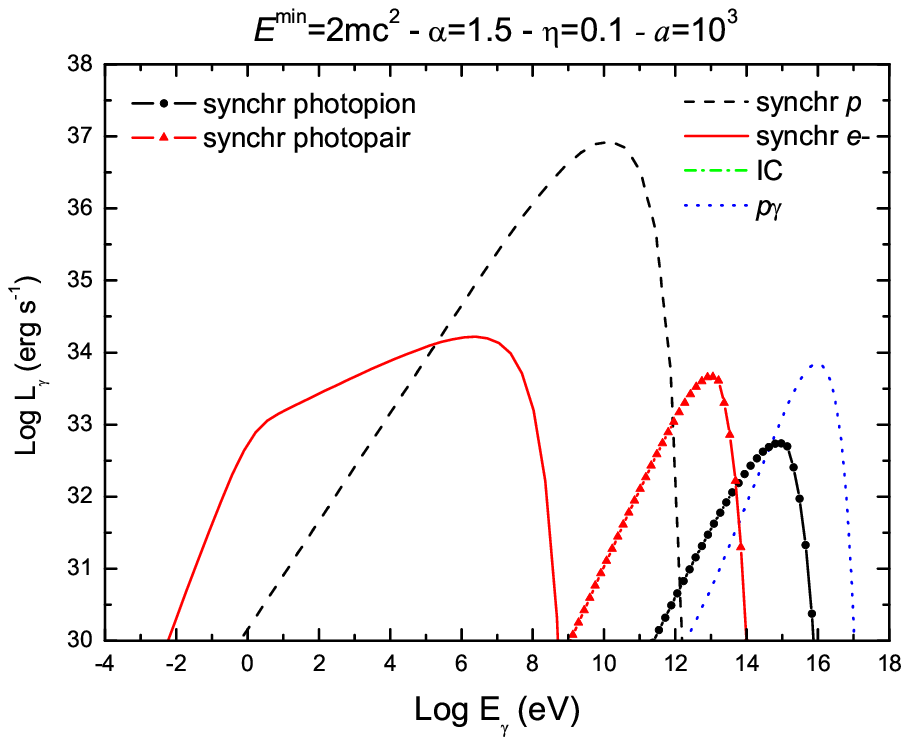} 
  \includegraphics[trim=10 18 5 0,clip,width=0.45\textwidth,keepaspectratio]{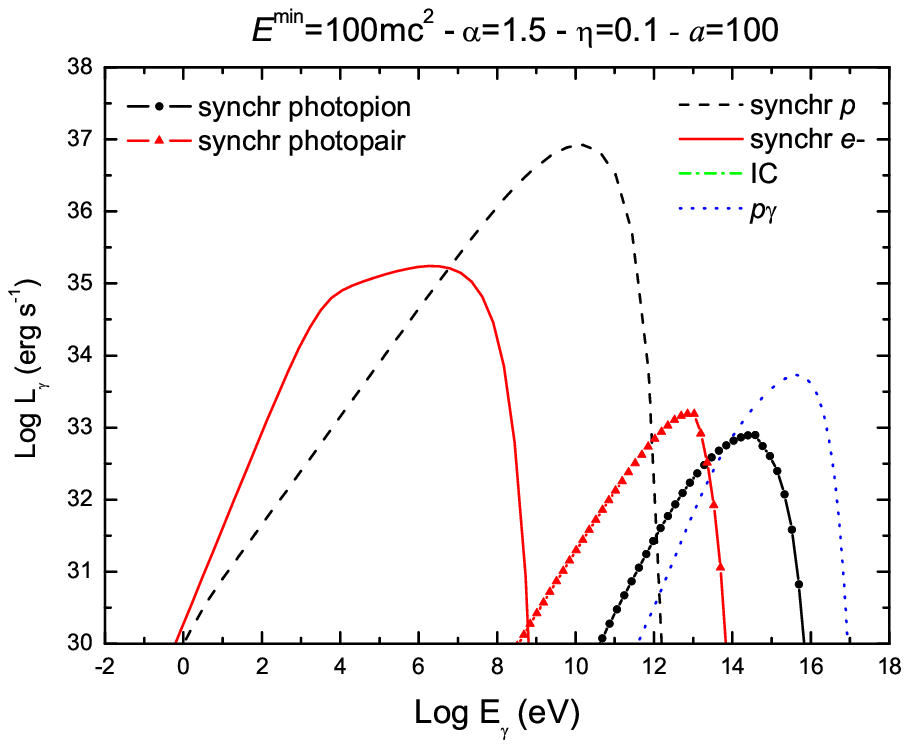} 
  \includegraphics[trim=10 18 5 18,clip,width=0.45\textwidth, keepaspectratio]{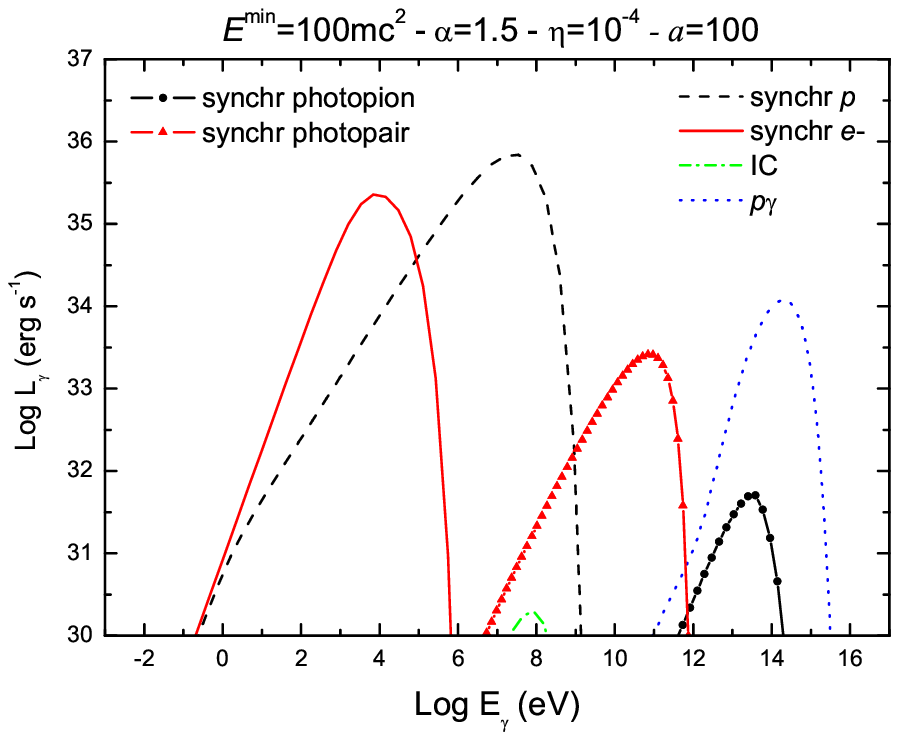}
  \caption{Spectral energy distributions obtained for different values of the parameters in the case of a particle injection index $\alpha=1.5$.}
  \label{fig:seds2}
\end{figure*}

Figures \ref{fig:seds1} and \ref{fig:seds2} show the spectral energy distributions (SEDs) calculated for different values of the model parameters. In Fig. \ref{fig:seds1} the spectral index of the injection function has the canonical value $\alpha=2.2$, whereas in Fig. \ref{fig:seds2} we considered a harder injection with $\alpha=1.5$. For each case, we show four representative spectra obtained by varying the value of the remaining parameters within a physically reasonable range. The main contributions to the SEDs are always due to synchrotron radiation of leptons and protons. The relative importance of each contribution depends on the value of the ratio $a=L_{p}/L_{e}$. The proton synchrotron spectra are hardly affected by changes in $a$, peaking at $\sim 10^{35-36}$ erg s$^{-1}$ for $E_\gamma\sim10^{8-9}$ eV; lepton synchrotron luminosities range from $\sim 10^{34}$ erg s$^{-1}$ for a completely proton-dominated jet ($a=10^3$) to $\sim 10^{37}$ erg s$^{-1}$ in the case of equipartition ($a=1$). In fact, lepton energy losses are so strong that they radiate almost all their available energy budget in each case.

The efficiency of the acceleration $\eta$ fixes the maximum particle energy and therefore has a direct effect on the high-energy cutoff of the SEDs. Synchrotron spectra extend up to $E_\gamma\sim10^{12}$ eV in the case of protons and up to $E_\gamma\sim10^{9}$ eV in the case of leptons for $\eta=0.1$; when a poor acceleration efficiency $\eta=10^{-4}$ is considered, only energies of about three orders of magnitude smaller are reached. In the same way, modifying the minimum particle energy accordingly changes the low-energy cutoff of the spectra. 

The IC contribution is always negligible and, for $a>1$, IC luminosities are below $\sim10^{30}$  erg s$^{-1}$ as the leptonic content of the jet is reduced. Photopion production ($p\gamma_{\pi^{0}}$) yields a very hard energy tail to the SEDs, peaking at energies $E_\gamma\sim10^{14-15}$ eV. The highest value of the $p\gamma_{\pi^{0}}$ luminosity is sensible to $a$, since the target photons for $p\gamma$ collisions are provided mostly by the lepton synchrotron radiation field. Luminosities as large as $10^{35-36}$ erg s$^{-1}$ can be reached at very high energies. Synchrotron radiation from $e^\pm$ produced in $p\gamma_{e^\pm, \pi^\pm}$ can be important in those models with relatively low barion content, as it can be seen from the different SEDs in Figs. \ref{fig:seds1} and \ref{fig:seds2}. In general, these contributions have luminosities not too different from those obtained from $p\gamma_{\pi^0}$ but they cover a lower energy range.

\section{Internal photon-photon attenuation}

Internal photon-photon absorption might be important in some of the models we have calculated. In particular, TeV gamma-rays can be efficiently absorbed by IR synchrotron photons. This might initiate an electromagnetic cascade in the emission region of those cases where there are significant photon fields at low energies as could happen in blazars (e.g., Blandford \& Levinson 1995). In the context of microquasars, however, a strong magnetic field can quench a pure IC cascade. This is the result of  synchrotron losses for the first few generations of pairs, which then cannot produce photons energetic enough to sustain the electromagnetic avalanche (Khangulyan et al. 2008). The photon absorption, nonetheless, might change the gamma-ray spectrum depending on the specific model. This, in turn, can yield a variety of slopes in the spectra to be observed by GLAST or other instruments sensitive in the MeV-up-to-TeV energy range (see, e.g., Aharonian et al. 2008 for a recent discussion in the context of extragalactic jets). 

The opacity of a photon field of density $n(\epsilon, z)$ given by Eq. (\ref{sylocal}) to the propagation of a gamma-ray photon of energy $E_{\gamma}$ is (Gould \& Schr\'eder 1966): 

\begin{eqnarray}
	\tau(E_{\gamma})=\frac{1}{2}\int_{l}\,\int^{\epsilon_{\rm max}}_{\epsilon_{\rm th}}\int^{u_{\rm max}}_{-1} (1-u)&& \sigma(E_{\gamma}, \epsilon, u)\nonumber\\ &&\times n(\epsilon, z)\; \mathrm{d}u \; \mathrm{d}\epsilon \;\mathrm{d}l,
\end{eqnarray}

\noindent where $u=cos \vartheta$, $\vartheta$ is the angle between the momenta of the colliding photons, $l$ is the photon path, and the cross section for the interaction is given by (e.g., Levinson 2006):
\begin{eqnarray}
	\sigma_{\gamma\gamma}(E_\gamma,\,\epsilon,&&\vartheta)=\frac{3}{16}\sigma_{\rm T}(1-\beta^2)\nonumber\\&&\times\left[\left(3-\beta^4\right)\ln\left(\frac{1+\beta}{1-\beta}\right)-2\beta\left(2-\beta^2\right)\right].
\end{eqnarray}
In the latter expression, $\beta$ is the speed of the electron/positron in the center of momentum frame. It is related to the energy of the incident and target photons by:

\begin{equation}
(1-\beta^2)=\frac{2m_e^2c^4}{(1-u)E_\gamma\epsilon}; \qquad 0\leq\beta<1.	
\end{equation}
The threshold energy $\epsilon_{\rm th}$ is defined by the condition $\beta=1$ with $\vartheta=0$. 

\begin{figure}[!h]
  \centering
  \includegraphics[trim=20 15 5 30,clip,width=0.45\textwidth, keepaspectratio]{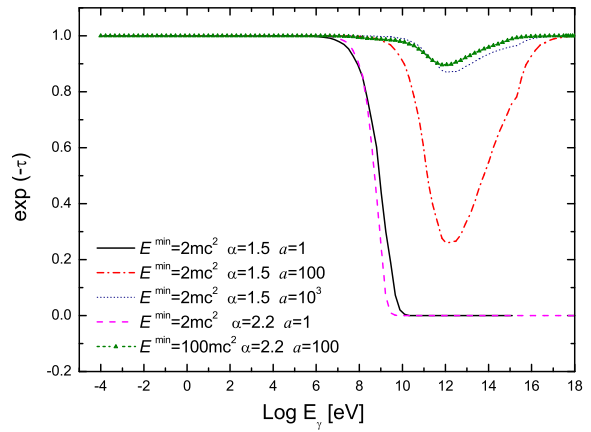}
  \caption{Attenuation curves for different models of a proton microquasar. In all cases the acceleration efficiency is $\eta=0.1$.}
  \label{fig:atten}
\end{figure}

We have calculated the attenuation factor $\exp(-\tau)$ for the different models presented in the previous section. Some results are shown in Fig. \ref{fig:atten}. 

We notice that in all cases with significant leptonic synchrotron emission, the radiation is completely suppressed above 10 GeV. For proton-dominated models, which are characterized  by a prominent proton synchrotron peak, the attenuation is quite moderate. 

\begin{figure*}[!t]
  \centering
  \includegraphics[trim=20 18 5 0,clip,width=0.45\textwidth, keepaspectratio]{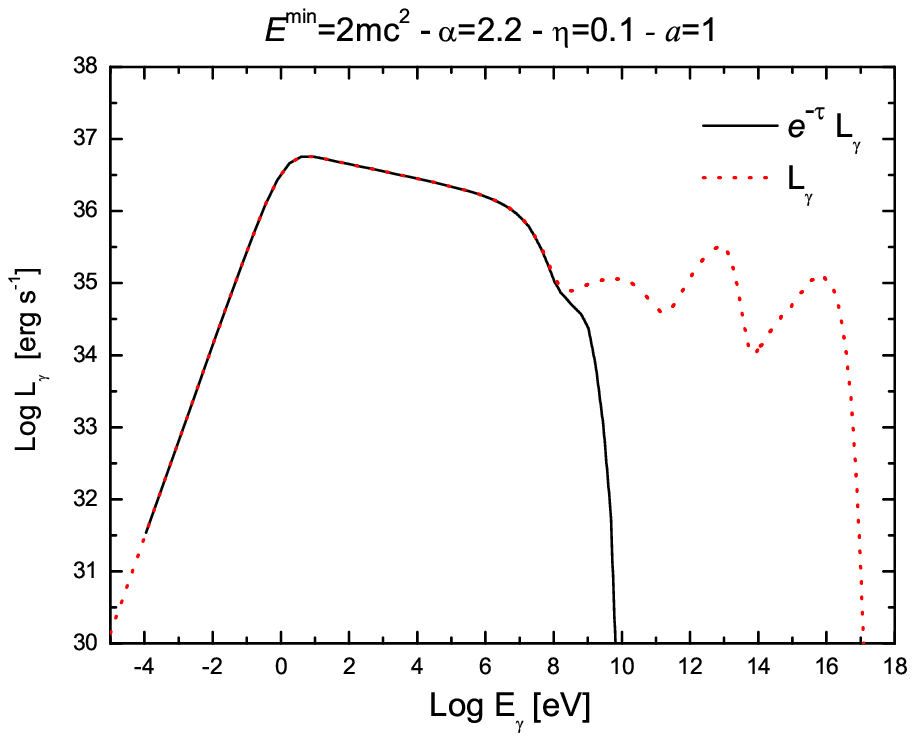}
  \includegraphics[trim=20 18 5 0,clip,width=0.45\textwidth, keepaspectratio]{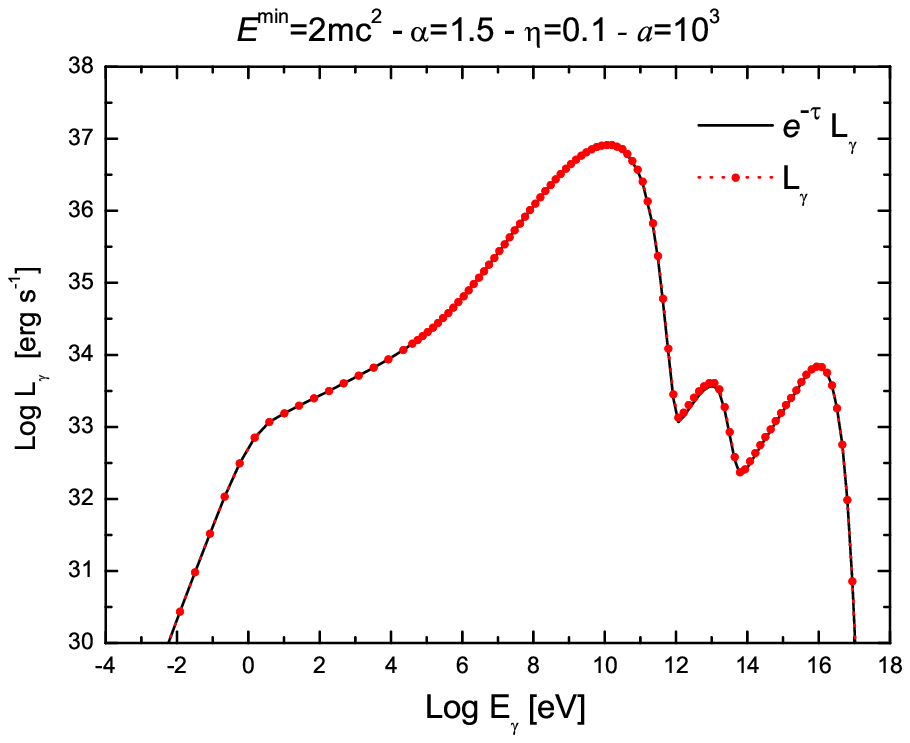} 
  \caption{Spectral energy distributions attenuated by internal absorption for two different microquasar models. Notice that in the case $a=1$ the emission above $\sim 10 $ GeV is completely suppressed. On the contrary, for $a=10^{3}$ the production SED is basically unmodified.}
  \label{SEDabsorb}
\end{figure*}

In Fig. \ref{SEDabsorb},  we present two spectral energy distributions modified by the effects of the photon-photon absorption. In the left panel, we show a case with $a=1$ (strong attenuation). The internal opacity to gamma-ray propagation results in a significant softening of the spectrum between $\sim10$ MeV and $\sim10$ GeV, with a cutoff beyond the latter energy. This is consistent with the type of very soft spectra observed by EGRET in some variable halo sources (e.g., Thompson et al. 2000, Grenier 2001). In the right panel of the same figure, we show the SED for a proton-dominated microquasar, which is basically not modified by absorption effects.  

In all cases the emission at TeV energies is either suppressed or too weak, and hence no detection with current Cherenkov telescope arrays should be expected for these objects unless very high acceleration efficiencies could be achieved in proton-dominated cases\footnote{A high acceleration efficiency would move the high-energy cutoff of the synchrotron peak into the TeV regime.}. Neutrino propagation is not affected by the effects discussed in this section.

\section{Discussion}

From the results presented in the previous sections is clear that proton low-mass microquasars can be significant gamma-ray sources in the MeV-GeV range, and perhaps in some cases even at TeV energies. According to the ratio $a$ of primary protons to leptons, we can obtain different types of low-energy counterparts. Such counterparts range from sources with radio luminosities of $10^{29}-10^{30}$ erg s$^{-1}$ and strong X-ray emission (e.g., the model shown in the left upper panel of Fig. \ref{fig:seds2}), up to sources with weak luminosities dominated by proton synchrotron emission between 100 MeV and 100 GeV (e.g., the model shown in the right upper panel of Fig. \ref{fig:seds2}). Since the cutoff of the synchrotron peak always occurs at energies below 1 TeV, and the high-energy emission due to photopion production is beyond 10 TeV in all models, detection with ground-based Cherenkov telescopes with a threshold of around 1 TeV would be difficult. In cases with strong leptonic synchrotron emission, the TeV radiation is directly suppressed by internal absorption.  The telescope GLAST, on the contrary, with its energy window of 100 MeV $-$ 300 GeV is especially suitable for the detection of these objects. Actually, we suggest that many of the unidentified EGRET sources detected off the galactic plane around the Galactic center, might be proton microquasars. Variability, one of the outstanding properties of these sources (Nolan et al. 2003), can be easily introduced in our model through a variable accretion rate, precessing jets, or internal shocks (in this latter case, there would be very rapid variability, superposed to longer variations with timescales from hours to days). The Astro-rivelatore Gamma a Immagini LEggero (AGILE) satellite might provide a more accurate location for some of these sources in the immediate future. Low-threshold, ground-based gamma-ray Cherenkov telescope arrays like HESS II and MAGIC II could also detect hadronic LMMQs at $E\geq100$ GeV. In particular, the tail of the proton synchrotron peak from models with large values of $a$ might be detectable, displaying a soft spectra. So, observations of LMMQs with Cherenkov telescopes can be useful to constrain the parameter $a$. 

\vspace{-0.04cm}

An additional prediction of our model is the production of high-energy ($E>1$ TeV) neutrinos, with luminosities in the range $10^{33}-10^{35}$ erg s$^{-1}$. In the cases of the highest luminosities, if the MQ is located not too far away (say, around 2 kpc as it is the case of XTE J1118+480) the expected flux could be similar to those estimated for HMMQs (e.g., Levinson \& Waxman 2001; Romero et al. 2003; Christiansen et al. 2006; Aharonian et al. 2006). We notice that heavily absorbed sources are precisely those with the highest neutrino luminosities, so a correlation between gamma-ray and neutrino fluxes should not necessarily be expected. Finally, we mention that some features related to electron/positron annihilation could be observable in cases with strong absorption (e.g., Romero 1996; Punsly et al. 2000; Levinson 2006).  

\section{Conclusions} 

We have developed a simple model for proton low-mass microquasars that predicts significant gamma-ray emission in the MeV and GeV domain for these objects. Observations with GLAST will help to establish some fundamental parameters such as the ratio of relativistic protons to electrons in the sources and the effects of internal absorption. Proton low-mass microquasars also produce very high-energy gamma-rays (well within the TeV range) and neutrinos, that in some cases might be detectable by km$^{3}$-detectors like IceCube or Nemo. Proton microquasars might also present very weak low energy counterparts, and hence they are suitable candidates to explain the unidentified gamma-ray sources detected by EGRET in the galactic halo. 

Future developments of the model should include a more sophisticated treatment of the acceleration region and time-dependent calculations. These topics will be included in a forthcoming communication.

\begin{acknowledgements}
We are grateful to an anonymous referee for constructive remarks on the manuscript. We thank Prof. S. R. Kelner for useful comments on radiative processes. This work has been supported by grants PIP 5375 (CONICET) and PICT 03-13291 
BID 1728/OC-AR (ANPCyT). GER acknowledges support by the Ministerio de Educaci\'on y Ciencia (Spain) under grant AYA 2007-68034-C03-01, FEDER funds.
\end{acknowledgements}


\begin{thebibliography}{99}
\bibitem{}Aharonian, F.A., \& Atoyan, A.M. 1998, New Astron. Rev., 42, 579
\bibitem{}Aharonian, F.A. 2002, MNRAS, 332, 215
\bibitem{}Aharonian, F. A., et al. (HESS Coll.) 2005, Science, 309, 746
\bibitem{}Aharonian, F. A., et al. 2006, J. Phys. Conf. Ser., 39, 408
\bibitem{}Aharonian, F.A., Khangulyan, D., \& Costamante, L. 2008, MNRAS, submitted [arXiv:0801.31198v1]
\bibitem{}Albert, J. et al. (MAGIC coll.) 2006, Science, 312, 1771
\bibitem{}Albert, J. et al. (MAGIC coll.) 2007, ApJ, 665, L51
\bibitem{}Atoyan, A.M., \& Aharonian, F.A. 1999, MNRAS, 302, 253
\bibitem{}Atoyan, A.M., \& Dermer, C.D. 2003, ApJ, 586, 79
\bibitem{}Blandford, R.D., \& Levinson, A. 1995, ApJ, 441, 79
\bibitem{}Blumenthal, G.R., \& Gould, R.J. 1970, Rev. Mod. Phys., 42, 237
\bibitem{}Bosch-Ramon, V., Romero, G.E., \& Paredes, J.M. 2005, A\&A, 429, 267
\bibitem{}Bosch-Ramon, V., Romero, G.E., \& Paredes, J.M. 2006, A\&A, 447, 263
\bibitem{}Bosch-Ramon, V. 2007, Ap\&SS, 309, 321
\bibitem{}Chodorowski, M.J., Zdziarski, A.A., \& Sikora, M. 1992, ApJ, 400, 181
\bibitem{}Christiansen, H.R., Orellana, M., \& Romero, G.E. 2006, Phys. Rev. D., 73, 063012
\bibitem{FB}Falcke, H., \& Biermann, P. 1995, A\&A, 293, 665
\bibitem{}Ghisellini, G., Maraschi, L., \& Treves, A. 1985, A\&A, 146, 204
\bibitem{}Ginzburg, V.L., \& Syrovatskii, S.I. 1964, The Origin of Cosmic Rays, Pergamon Press, Oxford
\bibitem{}Gould, R.J., \& Schr\'eder, G.P. 1996, Phys. Rev., 155, 1404 
\bibitem{}Grenier, I.A. 2001, in: A. Carrami\~nana et al. (eds.), The Nature of Unidentified High-Energy Gamma-Ray Sources, Kluwer Academic Publishers, Dordrecht, p. 51
\bibitem{}Grenier, I.A. 2004, in: K.S. Cheng \& G.E. Romero (eds.), Cosmic Gamma-Ray Sources, Kluwer Academic Publishers, Dordrecht, p. 47
\bibitem{}Grenier, I.A., Kaufman Bernad\'o, M.M., \& Romero, G.E. 2005, Ap\&SS, 297, 109 
\bibitem{}Kaufman Bernad\'o, M.M., Romero, G.E., \& Mirabel, I.F. 2002, A\&A, 385, L10
\bibitem{}Khangulyan, D., Hnatic, S., Aharonian, F.A., \& Bogovalov, S.  2007, MNRAS, 380, 320
\bibitem{}Khangulyan, D., Aharonian, F.A. \& Bosch-Ramon, V. 2008, MNRAS, 383, 467
\bibitem{Kord} K\"ording, E.G., Fender, R.P., \& Migliari, S. 2006, MNRAS, 369, 1451
\bibitem{}Levinson, A., \& Blandford, R.D. 1996, ApJ, 456, L29
\bibitem{}Levinson, A., \& Waxman E. 2001, Phys. Rev. Lett., 87, 171101
\bibitem{}Levinson, A. 2006, Int. J. Mod. Phys. A, 21, 6115
\bibitem{}Lind, K.R., \& Blandford, R.D. 1985, ApJ, 295, 358
\bibitem{}Markoff, S., Falcke, H., \& Fender, R.P. 2001, A\&A, 372, L25
\bibitem{}Mastichiadis, A., Protheroe, R.J., \& Kirk, J.G. 2005, A\&A, 433, 765
\bibitem{}Maximon, L.C. 1968, J. Res. NBS, 72B, 79
\bibitem{}Migliari, S., \& Fender, R.P. 2006, MNRAS, 366, 79
\bibitem{M1}Mirabel, I.F., Rodr\'\i guez, L.F., Cordier, B., Paul, J., \& Lebrun, F. 1992, Nature, 358, 215
\bibitem{M4}Mirabel, I.F. \& Rodr\'{\i}guez, L.F. 1999, ARA\&A, 37, 409
\bibitem{}M\"ucke, A., Engel, R., Rachen, J.P. et al. 2000, Comm. Phys. Comp., 124, 290
\bibitem{}Nolan, P.L, Tompkins, W.F., Grenier, I.A., \& Michelson, P.F. 2003, ApJ, 597, 615 
\bibitem{}Punsly, B., Romero, G.E., Torres, D.F., \& Combi, J.A. 2000, A\&A, 364, 552
\bibitem{}Romero, G.E. 1996, A\&A, 313, 759
\bibitem{}Romero, G.E. 2001, in: A. Carrami\~nana et al. (eds.), The Nature of Unidentified High-Energy Gamma-Ray Sources, Kluwer Academic Publishers, Dordrecht, p. 65
\bibitem{}Romero, G.E., Torres, D.F., Kaufman Bernad\'o M.M., \& Mirabel I.F. 2003, A\&A, 410, L1
\bibitem{}Romero, G.E. 2005, Chinese J. Astron. \& Astrophys., 5, 110
\bibitem{}Romero, G.E., Okazaki, A.T., Orellana, M., \& Owocki, S.P. 2007, A\&A, 474,15
\bibitem{}Thompson, D.J., Bertsch, D.L., Hartmann, R.C., et al. 2000, in: M.L. McConnell and J.M. Ryan (eds.), AIP Conference Proceedings, 510, 479 


\end{thebibliography}
\end{document}